\documentclass{article}

 \usepackage[preprint]{neurips_2022}

\usepackage{natbib}
\setcitestyle{numbers, compress}
\usepackage[utf8]{inputenc} 
\usepackage[T1]{fontenc}    
\usepackage{hyperref}       
\usepackage{url}            
\usepackage{booktabs}       
\usepackage{amsfonts}       
\usepackage{nicefrac}       
\usepackage{microtype}      
\usepackage{xcolor}         
\usepackage{graphicx}
\usepackage{amsmath}
\usepackage{nicefrac}
\usepackage{fdsymbol}
\usepackage{caption}
\usepackage{subcaption}

\title{Motion ID: Human Authentication Approach}

\author{%
  Aleksei Gavron\\
  Samsung R\&D Institute Rus\\
  Russia, Moscow, 127018\\
  \texttt{a.gavron@samsung.com}\\
  \And
  Konstantin Belev\\
  Samsung R\&D Institute Rus\\
  Russia, Moscow, 127018\\
  \texttt{k.belev@samsung.com}\\
  \And
  Konstantin Kudelkin\\
  Samsung R\&D Institute Rus\\
  Russia, Moscow, 127018\\
  \texttt{k.kudelkin@samsung.com}\\
  \AND
  Vladislav Shikhov\\
  Samsung R\&D Institute Rus\\
  Russia, Moscow, 127018\\
  \texttt{v.shikhov@samsung.com}\\
  \And
  Andrey Akushevich\\
  Samsung R\&D Institute Rus\\
  Russia, Moscow, 127018\\
  \texttt{a.akushevich@samsung.com}\\
  \And
  Alexey Fartukov\\
  Samsung R\&D Institute Rus\\
  Russia, Moscow, 127018\\
  \texttt{a.fartukov@samsung.com}\\
  \AND
  Vladimir Paramonov\\
  Samsung R\&D Institute Rus\\
  Russia, Moscow, 127018\\
  \texttt{v.paramonov@samsung.com}\\
  \And
  Dmitry Syromolotov\\
  Samsung R\&D Institute Rus\\
  Russia, Moscow, 127018\\
  \texttt{d.syromoloto@samsung.com}\\
  \And
  Artem Makoyan\\
  Samsung R\&D Institute Rus\\
  Russia, Moscow, 127018\\
  \texttt{a.makoyan@samsung.com}\\
}

\begin{document}

\maketitle

\begin{abstract}
We introduce a novel approach to user authentication called Motion ID. The method employs motion sensing provided by inertial measurement units (IMUs), using it to verify the person's identity via short time series of IMU data captured by the mobile device. The paper presents two labeled datasets with unlock events: the first features IMU measurements, provided by six users who continuously collected data on six different smartphones for a period of 12 weeks. The second one contains 50 hours of IMU data for one specific motion pattern, provided by 101 users. Moreover, we present a two-stage user authentication process that employs motion pattern identification and user verification and is based on data preprocessing and machine learning. The Results section details the assessment of the method proposed, comparing it with existing biometric authentication methods and the Android biometric standard. The method has demonstrated high accuracy, indicating that it could be successfully used in combination with existing methods. Furthermore, the method exhibits significant promise as a standalone solution. We provide the datasets to the scholarly community and share our project code.
\end{abstract}

\section{Introduction}
\label{sec:intro}

Traditional authentication methods in mobile biometrics (PIN or lock pattern) are being phased out in favor of modern approaches such as fingerprint\cite{maltoni2009handbook}, iris\cite{odinokikh2018high}, or facial recognition\cite{wang2021deep}. The most important benefits of such a switch are user convenience and greater recognition accuracy. However, while the latest technologies have their strengths, both types of biometric methods entail additional interaction between the user and the device, thus being called explicit authentication. This interaction brings along a whole array of issues.

The first issue lies in the necessity of additional hardware in mobile devices. Fingerprint recognition systems require optical, capacitive, or ultrasonic\cite{strohmann2020ultrasonic} sensors. For iris recognition, smartphones must be equipped with special infrared cameras. Due to the emergence of new security issues and corner cases (which we will discuss below), facial recognition needs special equipment, such as a time-of-flight camera (e.g. TrueDepth) or a stereo camera, rather than just a regular selfie camera. Mobile hardware is constantly changing to keep up with the trends - for example, fingerprint sensors and cameras are being integrated underneath the displays. This process is extremely costly. Furthermore, hardware changes require changes in the software, which only worsens the issue at hand.

The second issue is insufficient security. Fingerprint, iris, and face recognition systems can be hijacked via artificial fingerprints/irises/faces (spoofs), manually manufactured through different methods: silicone or gelatin fakes (for fingerprints), 2D photos, or printed 3D masks (for iris and face recognition). Currently, specific anti-spoofing algorithms\cite{odinokikh2018iris,nogueira2016fingerprint} are used to tackle this issue. But anti-spoofing algorithms have to be constantly updated to successfully counteract the use of different kinds of lighting, materials, and the general advances in hacking techniques, making it a never-ending endeavor. This is a widespread weakness that has led to the emergence of international anti-spoofing competitions such as LivDet\cite{casula2021livdet,8272763}.

The third issue concerns the emergence of new corner cases. During the pandemic, masks\cite{wang2020masked} and gloves have become a significant obstacle to facial recognition and fingerprint recognition systems, respectively. This lowers the accuracy of facial recognition, requiring additional tweaks and updates to software or hardware, while fingerprint recognition systems become unusable.

The fourth issue concerns the limited application of such methods. As far as mobile business is concerned, all the above-mentioned biometric technologies are used only in smartphones. However, more and more different devices require biometrics. This includes AR/VR glasses, gaming controllers, smartwatches and other wearable electronics. Not all of these devices can support built-in cameras or fingerprint sensors, to say nothing of ease of use. In addition, the emergence and popularity of metaverses\cite{lee2021all} mean that additional hardware for user authentication is becoming a must.

To sum up, the above-mentioned issues prevent manufacturers from fully complying with the requirements for biometric technology, presented in the Android Compatibility Definition Document.\footnote{CDD: (\url{https://source.android.com/security/biometric/measure})} Passive (or implicit) authentication\cite{fartukov2021mobile}, based on IMU sensors, can circumvent these issues without sacrificing security and ease of use\cite{deb2019actions}. Widespread usage of IMU sensors in smartphones and other wearable electronics allows us to create a unique system that tracks and recognizes motion patterns for each specific user.

\section{Related Work}

There are very few datasets with recorded IMU data and they are all different in a variety of ways. This indicates that there are no standards for collection of such data, which further complicates the search for suitable data for our experiment.

Data for UCI-HAR\cite{anguita2013public} were collected by 30 users, with each user performing six actions: walking, walking upstairs and downstairs, sitting, standing, and lying down. Users also wore their smartphones at the waist level. This dataset was unable to solve the previously mentioned biometrics problems for several reasons: users did not use the smartphone in the way that people would ordinarily use, the dataset did not capture unlocking events, and the number of users was too low to enable optimal performance that is required for biometrics. The HMOG dataset\cite{yang2014multimodal} contains a much larger amount of IMU data - 100 volunteers collected data over randomly selected actions, such as reading, writing, or map navigation. However, this dataset does not fully cover all aspects of the problem either, most importantly, it lacks flag labels for unlock events. The WISDM-HARB dataset\cite{weiss2019smartphone} was collected from 51 participants during 18 different activities. IMU data were recorded at a sampling rate of 20 Hz via a smartphone and a smartwatch. This sampling rate is insufficient for the purposes of mobile user verification. Similar to the previous dataset, this one lacked labels for unlock events.

There is also a large-scale study\cite{neverova2016learning} on biometric authentication that relied on a huge (but private) dataset collected by 1,500 volunteers. Besides IMU measurements, this dataset also contains such smartphone sensor readings as images from the front camera, touchscreen data, GPS, Bluetooth, etc. However, tracking this amount of data is impractical because the resources available to biometric authentication systems are extremely limited and it is currently impossible to store all of the necessary data. There is a study\cite{deb2019actions} that takes such hardware limitations into account. It uses only the most relevant data that are most frequently mentioned in scholarly literature. However, the dataset for this study was collected by 30 volunteers with a sampling rate of 1 Hz. This is insufficient for identifying possible unlocking events and staying within the current performance requirements for biometrics.

\section{Motion ID: Concept of Operations}

Motion ID starts with a built-in pre-trained base algorithm based on two sets of data. For a certain period (a couple of weeks), the smartphone collects the owner's IMU data. The collected data is flagged at points when the user unlocks their phone. The system then adapts to the unique user data and the fine-tuned Motion ID system is ready for daily use.

\begin{figure}
  \centering
  \includegraphics[width=13cm]{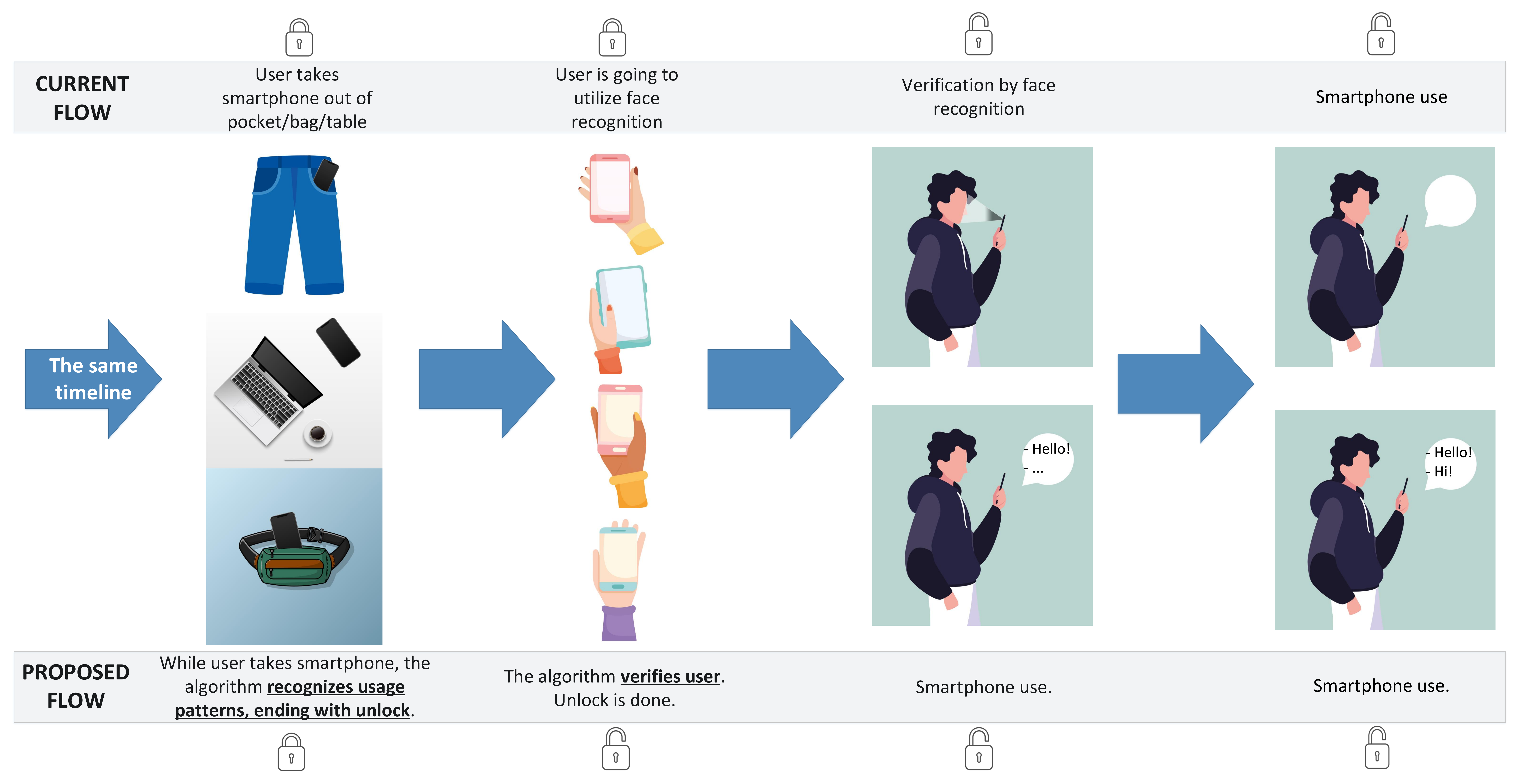}
  \caption{User scenario for Motion ID. The closed padlock icon (at the top and bottom) signifies the stages where the device has not yet been unlocked by the user, and, conversely, the open padlock icon denotes stages when the device is unlocked.}
  \label{fig:scenario}
\end{figure}

Once Motion ID is configured for a specific user, authentication occurs in two consecutive steps: (1) \emph{the pre-unlocking step}; (2) \emph{user verification} (to make sure that it was the owner who performed the action that fits the motion pattern). In the first step, the system predicts that the user is going to unlock the device by detecting a \emph{usage pattern} that corresponds to the device being unlocked. Next, the system verifies that it was the owner of the device who executed the pattern. This is done to rule out hijacking attempts (which would replicate the usage pattern). In other words, \emph{user verification} is performed.

Figure \ref{fig:scenario} shows a real-life user scenario for Motion ID (compared to facial recognition as an example). In the first step, when the owner is about to use their device, the proposed authentication method \emph{predicts that the unlock event will happen based on the usage patterns}. The system \emph{verifies the user} just before the owner is ready to use the device. The proposed method not only eliminates all existing obstacles for biometrics, but also reduces the time it takes to authenticate and verify the user (excluding the inference time of the biometric solution itself).

\section{IMU datasets}
\label{sec:dataset}

\subsection{Data Collection}
We have developed two smartphone data collection protocols and an Android app to collect IMU data. These protocols are based on two different types of interaction between the user and the smartphone, both of which are an integral part of Motion ID: (1) The first dataset focuses on recognizing individual usage patterns. This is the first part of Motion ID, called \emph{Motion Patterns Identification}; (2) The second dataset, called \emph{User Verification}, follows directly from the previous one.

\subsubsection{Motion Patterns Identification}

Six users each with a Samsung Galaxy S10e smartphone collected IMU data every day for 2 weeks. At the end of the 2 weeks, the users switched smartphones with each other and restarted the process. Each user spent 2 weeks per smartphone during the whole data collection process, which took 12 weeks in total. Throughout the experiment, the Galaxy S10e was the main and only device of each user. The smartphones were used habitually and ordinarily, with the only difference from real-life scenarios being that the data collection app was always on.

Data were collected from the following sensors: accelerometer (gravity and linear acceleration), magnetometer, gyroscope, and rotation sensor. The sampling rate averaged at 50 Hz.

Each user unlocked their phone using biometrics for the entire duration of the data collection, namely fingerprint recognition. For each unlock event, data was labeled with a special flag. The flags were used to prepare the data so that the machine learning model could be trained on it. To ensure consistency in the data, each of the smartphones was unlocked only by fingerprint (via a capacitive sensor on the side panel of the device).

\subsubsection{User Verification}

For the user verification part, we used the same Android app, but only on one smartphone: Galaxy S20. The data was collected by 101 users, each of whom lifted the smartphone from the table 300 times, 50 times for each of the 6 locations of the device.

The data collection procedure for user verification was as follows: (1) The user lifts the locked smartphone from the table surface to a comfortable level; (2) The user unlocks the smartphone via an in-display ultrasonic fingerprint sensor; (3) The user locks the smartphone via Home button and puts the device; (4) The cycle is repeated for 50 times at each location, for a total of 300 times per person. Data collection usually took about 30 minutes per user. The test subjects were allowed to rest for 0.5-2 minutes between each location. This was necessary to keep their motion patterns natural and prevent mechanic motions, as well as to subdivide data into six clusters. Each unlock event was labeled with a special flag, same as for the \emph{Motion Patterns Identification}.

\subsection{Sample Size (User Count) Justification}
There were no special requirements for the number of users for the \emph{Motion Patterns Identification} stage. Each user was processed independently from the others. Only the duration of data collection matters. More IMU data means more samples that the machine learning model can be trained on. The following explanation applies only to the \emph{User Verification} part.

The main goal of the Motion ID feasibility study approach was to achieve performance comparable, to the Android biometrics standard – the Android Compatibility Definition Document (CDD) – and the existing biometric solutions, at least for a limited number of use cases. According to CDD, biometric systems can be divided into Class 3 (formerly \emph{Strong}), Class 2, (formerly \emph{Weak}), or Class 1 (formerly \emph{Convenience}). Strong class requires \emph{TAR\footnote{TAR -- True Acceptance Rate}(@FAR\footnote{FAR -- False Acceptance Rate}=1/50000)=90\%} metrics.

To determine the number of people required for data collection, we had to take into account the following factors: (1) Sufficient number of people/attempts to correctly evaluate the test dataset. In other words, the ability to produce the metrics described above (for a strong class); (2) Sufficient data for training and validation sets; (3) Reasonable number of subjects, considering limited conditions; (4) Reasonable amount of time for data collection per person.


Getting the right FAR metric requires a huge amount sample size and a lot of time to directly compare their performance. To estimate the sufficient number of people/tries, we decided to use the rule of 30, which states that “to be 90\% confident that the true error rate is within ±30\% of the observed error rate, there must be at least 30 errors”\cite{mansfield2006information,doddington2000nist}.

In the case of TAR(@FAR=1/50000)=90\% and 90\% confidence, this rule required us to make 300 genuine comparisons and 1.5M impostor comparisons. However, this rule assumes that the tests are independent (with different subjects randomly selected from the population). This means that, to fully comply with the rule, we needed at least 300 subjects for genuine comparisons and 1.5 million subjects for impostor comparisons, which is impossible. For performance evaluation, we decided to use the bootstrap method. Cross-comparison approach reduces the expected confidence level compared to the same number of independent comparisons.

\subsection{Datasets Pre-processing}
Before training the neural network and evaluating the results further, we undertook several pre-processing steps, independently for each of the two datasets.

\subsubsection{Motion Patterns Identification}
At this stage, we needed to get the time series that leads or does not lead to a device being unlocked. To determine the series that leads to an unlock, we looked at all the timestamps marked with the USER\_PRESENT flag.\footnote{\url{https://developer.android.com/reference/android/content/Intent}} For each of these timestamps, we gathered data from each sensor over a period of 3 seconds. Combining measurements from all sensors, we obtained the required time series. At times, the smartphone failed to take measurements for several seconds due to technical imperfections of the sensors. To obtain a fixed signal length, we only took those time series, where there were at least 100 readings for each sensor.

To determine the time series that did not lead to the device being unlocked, we first had to eliminate the times when the phone was motionless – i.e., when the linear accelerometer readings were zero for all 3 axes. After that, we took the time intervals between SCREEN\_OFF and the next SCREEN\_ON or USER\_PRESENT flag, discarding the last 3 seconds (since these 3 seconds lead to the device being unlocked).

\subsubsection{User Verification}

\begin{figure}
\centering
\begin{subfigure}{.5\textwidth}
  \centering
  \includegraphics[width=1.0\linewidth]{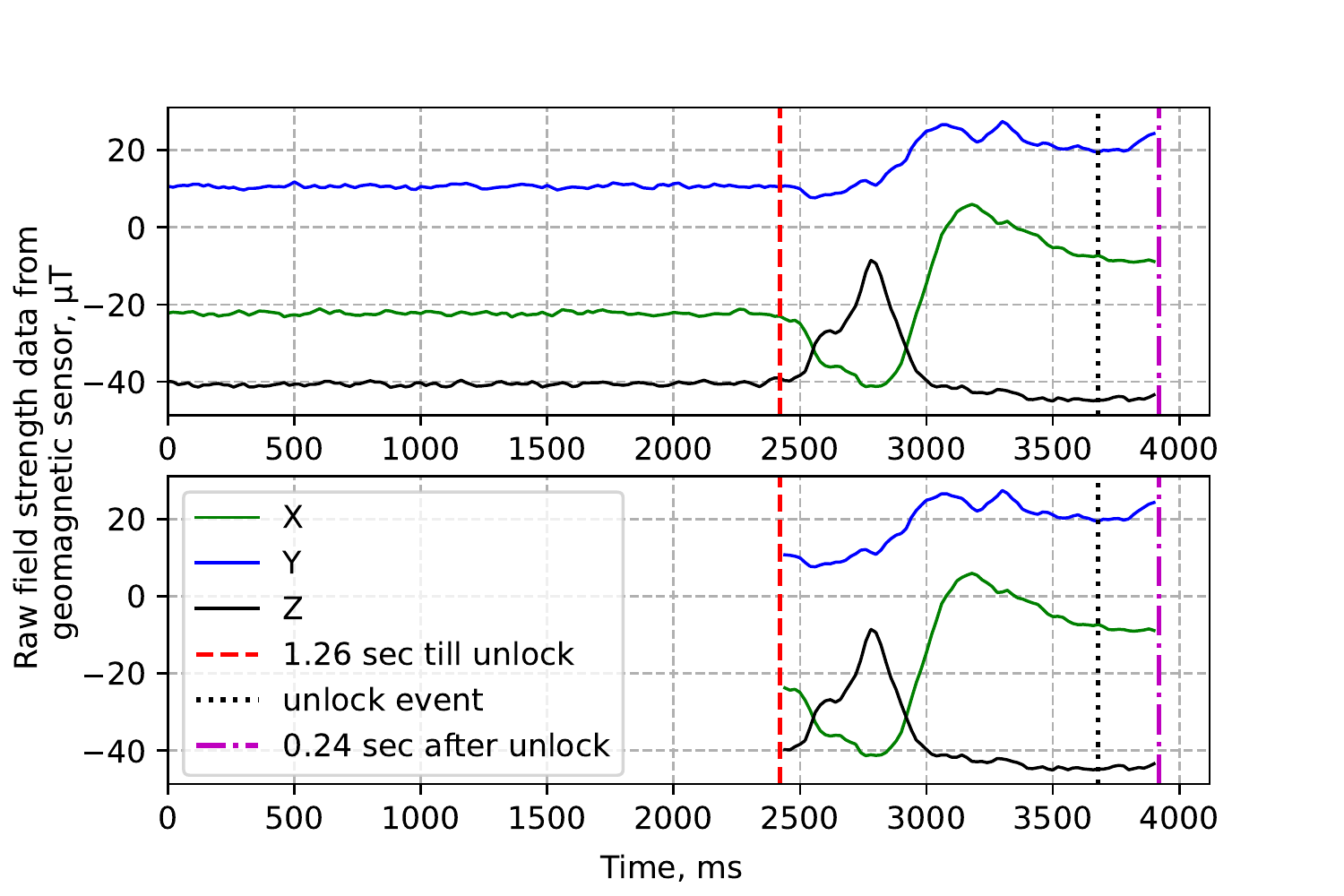}
  \label{fig:sub1}
\end{subfigure}%
\begin{subfigure}{.5\textwidth}
  \centering
  \includegraphics[width=1.0\linewidth]{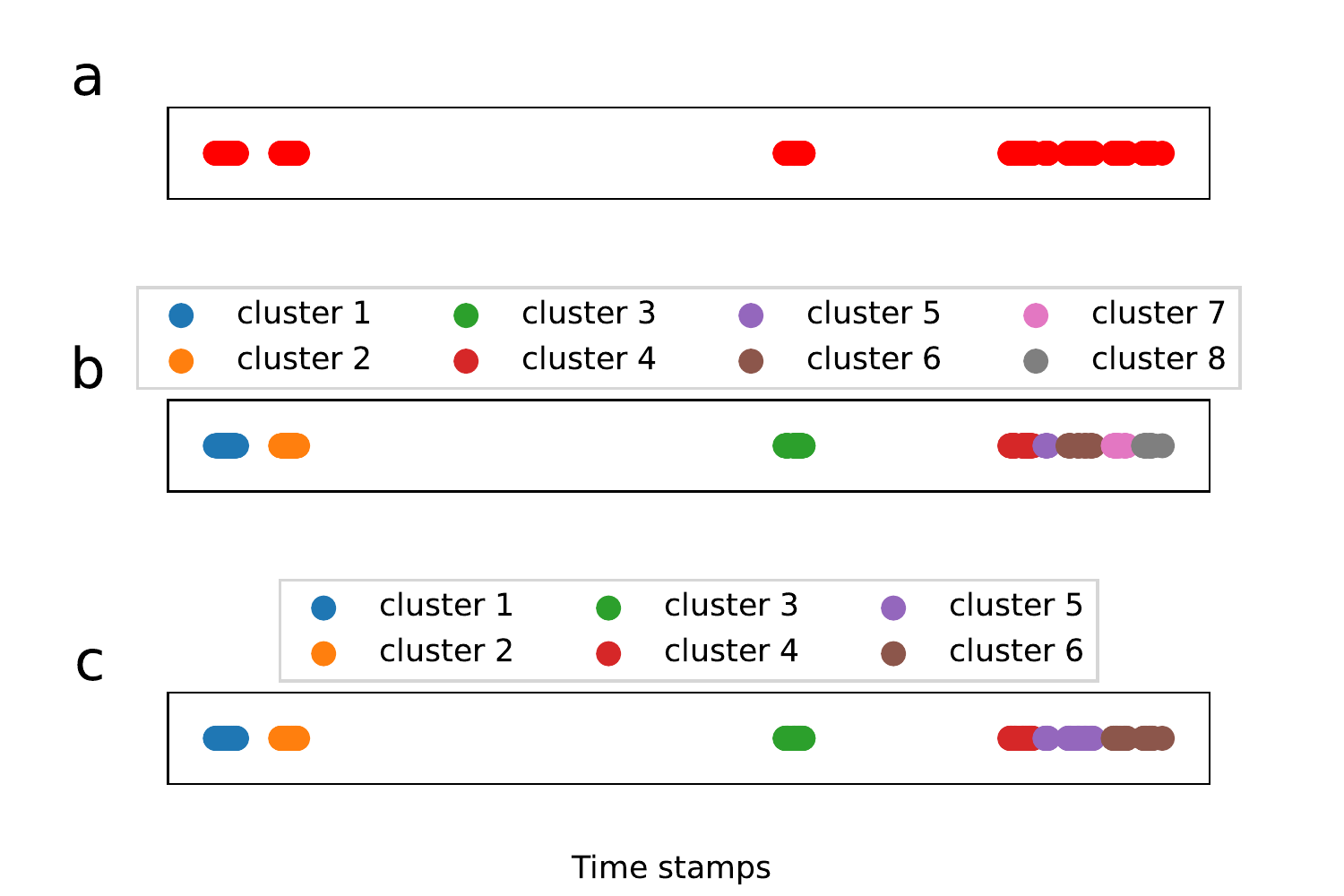}
  \label{fig:sub2}
\end{subfigure}
\caption{\emph{Left}: data collected by geomagnetic field sensor (\emph{x-axis} -- time, \emph{y-axis} -- raw field strength data in $\mu$T for X, Y, Z). \emph{Right}: subdividing the data into six clusters.}
\label{fig:magnetic_and_clusters}
\end{figure}


In this step, we collected data about the current state of the smartphone, subdivided into three states: (1) The smartphone is not in use; (2) The smartphone is currently in use; (3) The smartphone has just been unlocked.

The first pre-processing step is trimming the raw data. The data was collected constantly in all three states, with an average frequency of 50 Hz. However, we only need the readings that are directly before unlock event. As an example, Figure \ref{fig:magnetic_and_clusters} (left) shows the magnetometer data collected 4 seconds before unlocking, but we are only interested in the data from the last second for two reasons: 1) the most significant movements occur in the last second before unlocking; 2) inference time has a huge impact on biometrics in mobile devices. Therefore, we used only the data located between the red and magenta lines on the figure.

The second step was to subdivide the data into six clusters (Figure \ref{fig:magnetic_and_clusters} (right)) (corresponding to the six data collection locations) for each user. Here, ergonomics and the human factor played a more significant role than expected. The data collected at the first and sixth locations differed significantly. In the first location, subjects were only getting used to the motion pattern, the time between unlocking events was longer, the motion itself was unsteady, and subjects were adjusting to the size of the phone and its weight. Some of the users have never used in-display fingerprint biometry before, which took them some additional time to adjust to. By the end of data collection, most of these problematic aspects were gone.


\section{Motion Patterns Identification}
\label{sec:patterns}
\subsection{CNN architecture}
At this stage, the network has a traditional architecture consisting of several layers with pointwise convolutions, cross-entropy loss, and two output classes: true and false unlock events. The best epoch was selected according to the best ROC-AUC metric obtained on the validation set.

\section{User Verification} 
\label{sec:verification}
\subsection{Feature Generation} 
\label{subsec:features}
At this stage, the pre-processed data comprises accelerometer, rotation sensor, magnetometer, and gyroscope readings. The newly generated features are described below.

In smartphones, the accelerometer detects orientation changes and rotates the screen accordingly. In other words, the accelerometer readings use the screen as its frame. We converted accelerometer data to the Earth-fixed frame, which would determine the position of the smartphone in space. We rotate the readings from the gyroscope and magnetometer in a similar fashion.

The linear acceleration sensor outputs a three-dimensional vector representing acceleration along each device axis, excluding gravity. It is possible to directly obtain data from the linear accelerometer. But during the data collection, we noticed unstable sensor behavior: spikes in the data collection frequency, possibly caused by repetitive movements. We decided to manually generate measurements from the linear accelerometer and use it as another function. Theoretically, this sensor outputs readings according to the following formula (acceleration data, excluding acceleration due to gravity):
\[ linear\;acc = acc - acc\;due\;to\;gravity\]

The remaining generated features, which applies to almost all of the initial sensors (excluding initial raw data), consisted of: (1) Data rotated to the Earth-fixed frame; (2) Differences in measurements between the previous and the next reading (rotated and unrotated); (3) An integral of the sensor’s measurements (rotated and unrotated). The total number of feature vectors was 22.



\subsection{Data augmentation}
Time series of 1.5 seconds were randomly cut into segments of 1 second. Then, randomly distributed noise was added to each segment. Random noise serves as an additional regularization technique to prevent overfitting.

\subsection{Dataset splitting}
Implicit smartphone authentication systems utilize two main model training strategies: \emph{on-line} and \emph{off-line}\cite{deb2019actions}. Off-line approaches serve to train generic models that can be used to authenticate the user immediately after the application is installed and the user signs up. Therefore, the procedure of data set splitting is extremely important. The training, validation, and test sets must not have any overlaps in terms of users. In real-life conditions, the model is extremely likely to be overfitting, if it was not fine-tuned to a particular user. This means that the model would demonstrate adequate performance for users from the initial training dataset and far worse performance for the others. Within the confines of the proposed authentication method, the overfitting risk is exacerbated by the fact that distinguishing traits of the users are not entirely physiological, unlike irises or faces.

On-line approaches use a pre-trained baseline model that is further fine-tuned for each user individually. We employed this method in contrast with Android standard which sets metrics for off-line methods only. The on-line approach does not require splitting training, validation, and test datasets by users, so we subdivided datasets by attempts. According to the bootstrap method, we needed at least m = 188 attempts in the test dataset to theoretically estimate the metric defined in the Android biometric standard. This figure can be calculated by the formula: n*(n - 1)*m = 1.5M, where n -- number of users, m -- number of tries, 1.5M -- required number of impostor comparisons.

Below we will describe that, given the limitations of the trained model and the achieved performance, we can afford to use fewer users in the sample without a significant impact on the accuracy.

\subsection{CNN architecture}

\begin{figure}
  \centering
  \includegraphics[width=11cm]{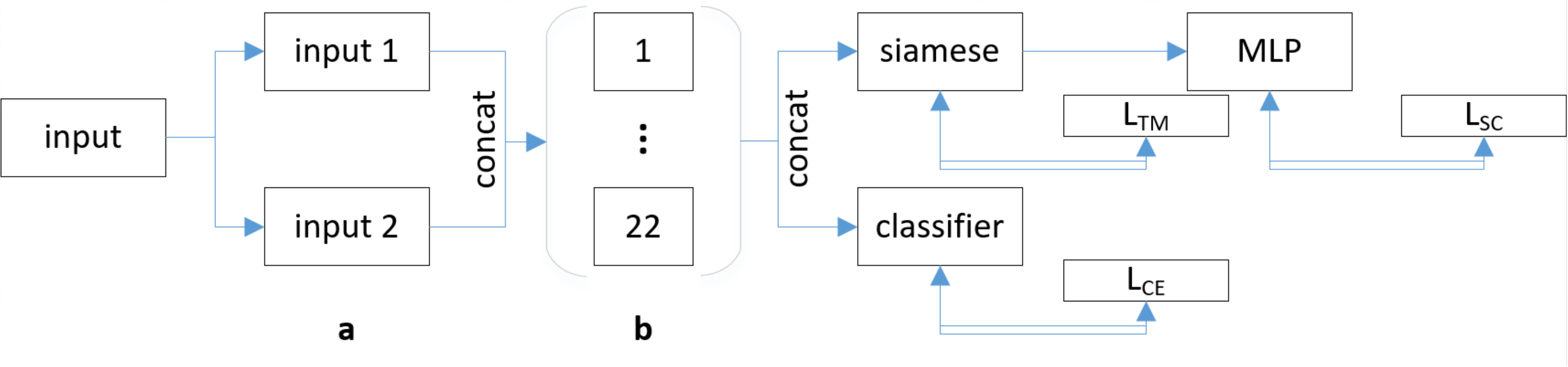}
  \caption{Block scheme of verification stage, where a) twice augmented input; b) 22 branches for each of the generated features; \emph{L}\textsubscript{CE} - Cross-Entropy loss; \emph{L}\textsubscript{TM} - Triplet Margin Loss, \emph{L}\textsubscript{SC} - Supervised Contrastive loss.}
  \label{fig:verification_pipeline}
\end{figure}

The entire architecture of the verification stage is depicted in Figure \ref{fig:verification_pipeline}. The input data (after feature generation) was augmented and concatenated twice. The first step is the feature extraction part, where separate CNN branches were created for each of the 22 generated data features (Figure \ref{fig:verification_pipeline}a). In other words, at this point, we wanted to get an information-rich conversion of the augmented raw data into an embedding space that could maintain the necessary distance between samples and, at the same time, be resistant to noise in the data. The input tensor was split into 22 equal parts, one part per feature. All branches have an identical architecture, consisting of 1D convolutional layers.

Further, the architecture splits into two. On one branch, the concatenated embeddings are processed via a classifier. The goal is to train the feature extractor in a way that allows its embeddings to contain information about the traits that are unique to each user. The target labels of this classifier are user IDs. Here, we employed cross-entropy loss as a loss function (Figure \ref{fig:verification_pipeline}, \emph{loss 1}) and defined it as:

\[ L_{\text CE} = -\sum_{i=1}^{n} t_i log(p_i), \] 
for n classes, where \emph{t}\textsubscript{i} -- the genuine user ID and \emph{p}\textsubscript{i} -- the Softmax function for the \emph{i}\textsuperscript{th} class.

The second branch begins with a Siamese\cite{koch2015siamese} network head, required for user verification. The goal of the Siamese head is to obtain highly discriminative features that can distinguish impostor comparisons from genuine ones. To train the Siamese head, we used the Triplet Margin Loss\cite{schroff2015facenet} function (Figure \ref{fig:verification_pipeline}, \emph{loss 2}). Let us assume that we are given three inputs: the anchor a, the positive p, and the negative n. The Triplet Margin Loss is defined as:

\[L_{\text TM}(a, p, n) = max\{d(a_i, p_i) - d(a_i, n_i) +\;margin, 0\} \]

where \[ d(x_i, y_i) = \lVert \textbf{x}_i - \textbf{y}_i \rVert_p\]

In the final step, embeddings from the Siamese head pass through a multi-layer perceptron. Usually, the cross-entropy function (the one we used in a previous classifier) is sensitive to adversarial examples and noise, outputting false negatives if the inputs differ from the initial data even just a little bit. IMUs demonstrate a significant hardware bias even in the case of repeated identical movements. To make the model more robust when fed augmented data, especially with added random noise, we applied a supervised contrastive pre-training method\cite{khosla2020supervised} to the classification task. The MLP head learns to map normalized embeddings of samples and their augmentations that belong to the same user closer to each other, and those belonging to any other user – farther.

Total loss is calculated in the following way:
\begin{align}
L_{\text total} = L_{\text CE} + \alpha_{\text TM} * L_{\text TM} + L_{\text SC}\nonumber
\end{align}

where $\alpha_{\text TM}$ - weighting coefficient for Triplet Margin loss.

\subsection{Training Procedure}
We propose the following training procedure and results evaluation method, consisting of four steps: 1) baseline model training; 2) user-specific fine-tuning for the trained model; 3) choosing the best epoch based on an additional validation subset; and 4) individual testing for each user from the final test subset. In steps 1–3, we took only 90 out of 101 users, the rest (11) were used in the final testing.

To train the baseline model, we used part of 90 users – i.e., \emph{n} users. These \emph{n} users were split into train\textsubscript{base}/val\textsubscript{base}/test\textsubscript{base} -- subset\textsubscript{base} by attempts (following the on-line approach). \emph{(90 - n)} users were used in \emph{additional validation} subset val\textsubscript{add}. We tried different split ratios for subset\textsubscript{base} and val\textsubscript{add}: from n = 60 to n = 85 with a step of 5. At this stage, the classifier in CNN architecture has n classes.

In the next stage, we fine-tune the trained baseline model for each of the 11 remaining users (test\textsubscript{final}). To do this, we a) use subset\textsubscript{base} (n users) and 11 users from test\textsubscript{final}; b) freeze feature extractor; c) reduce the learning rate and number of epochs; d) change number of classes in the classifier from n to 2, where the first class is user from subset\textsubscript{base} and the second class -- certain user from test\textsubscript{final}.

After the model is fine-tuned but before the final metrics are evaluated, we need to choose the suitable epoch correctly. For this, we used val\textsubscript{add}, mentioned above. Until this step, the \emph{90 - n} users from val\textsubscript{add} have not been used yet. As in the previous step, the classifier also has 2 classes, but in this case the first class is any user from val\textsubscript{add}, and the second class – certain user from test\textsubscript{final}. By testing 11 users, we selected an epoch with the best FAR\textsubscript{val}(@TAR=90\%) for each user.

The final test represents a real-world simulation of this approach: a pre-trained baseline model is fine-tuned to the device owner's data. Only the subset test\textsubscript{final} was used in the final testing. The first classifier also had 2 classes, where the first class is the current user and the second class -- any other user from test\textsubscript{final}. Here we used a bootstrap-like method, same as when training the baseline model. As usual, each user had 300 attempts. For testing, we used 90 of them. In addition to these 90 attempts by the current user, we also randomly chose 90 attempts by the remaining 10 users and repeated the point estimation of FAR 5000 times.

\section{Results}
\label{sec:results}

\begin{table}
  \caption{Motion Patterns Identification performance for each user-device pair}
  \label{table_patterns}
  \centering
  \begin{tabular}{lllllll}
    \toprule
    \multicolumn{7}{c}{Accuracy,\%}                   \\
    \cmidrule(r){2-7}
    Device/User    & 1     & 2     & 3     & 4     & 5     & 6\\
    \midrule
    1 & 85.5 ± 1.3  & 83.0 ± 1.2  & 79.4 ± 1.9  & N/A  & 84.4 ± 0.7  & 84.3 ± 1.2 \\
    2 & 88.7 ± 0.4  & 81.4 ± 1.5  & 82.1 ± 0.6  & 79.0 ± 2.0  & 83.7 ± 2.0  & 79.6 ± 1.2 \\
    3 & 91.0 ± 0.5  & 79.1 ± 1.1  & 73.6 ± 0.9  & 81.1 ± 4.0  & 81.7 ± 1.1  & 80.6 ± 1.2 \\
    4 & 82.2 ± 0.8  & 83.6 ± 0.2  & N/A  & 80.0 ± 2.0  & 87.6 ± 1.2  & N/A \\
    5 & 87.5 ± 1.0  & 79.1 ± 0.9  & 80.0 ± 0.7  & 80.9 ± 1.5  & 84.7 ± 1.5  & 82.2 ± 2.0 \\
    6 & 88.4 ± 0.5  & 77.6 ± 1.5  & 82.3 ± 0.9  & 82.0 ± 1.0  & 78.7 ± 1.3  & 79.1 ± 2.0 \\
    \bottomrule
  \end{tabular}
\end{table}

Table \ref{table_patterns} demonstrates Motion Patterns Identification accuracy for each user-device pair. Here, N/A denotes a lack of false unlock events or an insufficient amount thereof. The model predicted unlock events with fairly high accuracy. For a real-world application, these events must be continuously recorded and stored in memory. However, memory remains a limited resource for biometric algorithms, meaning that further studies in this area could focus on increasing the accuracy, reducing the size of the neural network, and optimizing data storage.

\begin{table}
  \renewcommand*{\arraystretch}{1}
  \caption{Validation and test performance for different splits}
  \label{table_ratios}
  \centering
  \begin{tabular}{lllllll}
    \toprule
    \multicolumn{6}{c}{Metrics of baseline model,\%}                   \\
    \cmidrule(r){2-6}
    split    & Acc\textsubscript{val}     & Acc\textsubscript{test}     & FAR\textsubscript{val}(@TAR=90\%)     & FAR\textsubscript{test}(@TAR=90\%)     & FAR\textsubscript{theor} \\
    \midrule
    60 & 97.9 ± 0.2  & 98.1 ± 0.3  & (1.0 ± 0.4)*10\textsuperscript{-2} & (2.0 ± 1.1)*10\textsuperscript{-2}  & 0.94*10\textsuperscript{-2} \\
    	& & & $1 \divslash 10000$ & $1 \divslash 5000$ &$1 \divslash 10620$ \\
	\cmidrule(r){1-6}
    65 & 98.2 ± 0.3  & 97.86 ± 0.16  & (0.5 ± 0.4)*10\textsuperscript{-2}  & (2.4 ± 0.5)*10\textsuperscript{-2}  & 0.8*10\textsuperscript{-2} \\
        & & & $1 \divslash 20000$ & $1 \divslash 4167$ &$1 \divslash 12480$ \\
        \cmidrule(r){1-6}
    70 & 97.9 ± 0.3  & 97.8 ± 0.2  & (1.0 ± 0.9)*10\textsuperscript{-2}   & (2.3 ± 0.2)*10\textsuperscript{-2}  & 0.69*10\textsuperscript{-2} \\
        & & & $1 \divslash 10000$ & $1 \divslash 4348$ &$1 \divslash 14490$ \\
        \cmidrule(r){1-6}
    75 & 98.1 ± 0.3  & 98.15 ± 0.15  & (0.8 ± 0.4)*10\textsuperscript{-2}  & (1.4 ± 0.6)*10\textsuperscript{-2}  & 0.6*10\textsuperscript{-2} \\
        & & & $1 \divslash 12500$ & $1 \divslash 7143$ &$1 \divslash 16650$ \\
        \cmidrule(r){1-6}
    80 & 98.12 ± 0.16  & 98.07 ± 0.18  & (0.5 ± 0.3)*10\textsuperscript{-2}  & (1.4 ± 0.6)*10\textsuperscript{-2}  & 0.53*10\textsuperscript{-2} \\
        & & & $1 \divslash 20000$ & $1 \divslash 7143$ &$1 \divslash 18960$ \\
        \cmidrule(r){1-6}
    85 & 97.9 ± 0.4  & 98.10 ± 0.11  & (0.9 ± 0.6)*10\textsuperscript{-2}  & (1.0 ± 0.5)*10\textsuperscript{-2}  & 0.47*10\textsuperscript{-2} \\
        & & & $1 \divslash 11111$ & $1 \divslash 10000$ &$1 \divslash 21420$ \\
    \bottomrule
  \end{tabular}
\end{table}

Table \ref{table_ratios} shows the training results of the baseline models for different splits of subset\textsubscript{base} and val\textsubscript{add}. In all experiments, we randomly sampled the attempts and trained each model 5 times. All metrics are calculated for val\textsubscript{base} and test\textsubscript{base}. All FAR values were obtained at fixed TAR = 90\%. FAR\textsubscript{theor} denotes the theoretically achievable FAR at each split. The second lines in FAR columns are FAR metrics presented in $1 \divslash k$ form, for easier comparison with the Android biometrics standard.

The more users go through training, the better the validation results, thanks to the increased number of genuine-impostor pairs. However, test results are less stable. The current architecture needs more generalization. This could be achieved with a smaller model capacity and more data. In any case, the results for the baseline model are already comparable to the current standard.

\begin{table}
  \renewcommand*{\arraystretch}{1}
  \caption{Test performance for user-split pairs}
  \label{table_tune}
  \centering
  \begin{tabular}{lllllll}
    \toprule
    \multicolumn{7}{c}{For each user id: FAR(@TAR=90),\%}                   \\
    \cmidrule(r){2-7}
    user id    & 60     & 65     & 70     & 75     & 80     & 85  \\
    \midrule
    0 & 0                   & 0               & 0                   & 0.6 ± 0.4     & 0                  & 1.0 ± 0.4 \\
    1 & 4.2 ± 2.0       & 6 ± 3         & 4 ± 3             & 2.0 ± 1.4     & 6 ± 5            & 6 ± 5 \\
    2 & 9 ± 3             & 12 ± 11     & 17 ± 9           & 10 ± 6         & 5 ± 4            & 8 ± 6 \\
    3 & 12 ± 2           & 17 ± 4       & 10 ± 2           & 12 ± 4        & 13 ± 7           & 14 ± 5 \\
    4 & 2.0 ± 0.9       & 1.0 ± 0.4   & 1.9 ± 1.6       & 5 ± 3          & 1.0 ± 0.4       & 8 ± 4 \\
    5 & 12 ± 3           & 14 ± 5       & 11 ± 5           & 11 ± 6         & 10 ± 3          & 10 ± 4 \\
    6 & 5 ± 3            & 3 ± 2          & 4 ± 3             & 2.8 ± 1.8    & 4 ± 3             & 4.7 ± 1.8 \\
    7 & 24 ± 4          & 23 ± 5        & 21 ± 5           & 22 ± 6        & 22 ± 4           & 22 ± 2 \\
    8 & 6 ± 2            & 8 ± 3          & 6 ± 3             & 5 ± 3          & 5 ± 2             & 4.9 ± 1.5 \\
    9 & 4 ± 3            & 2.0 ± 1.0    & 2.4 ± 1.4       & 4 ± 3          & 1.6 ± 0.6       & 2.2 ± 0.8 \\
    10 & 1.9 ± 1.1    & 0.5 ± 0.2    & 0.6 ± 0.4       & 0.6 ± 0.4    & 0                   & 0.5 ± 0.2 \\
    \bottomrule
  \end{tabular}
\end{table}

All trained baseline models were used for testing each user from test\textsubscript{final}. Results are shown in Table \ref{table_tune}. The results obtained are highly dependent on the users themselves. This is most likely caused by user-specific traits in hand motions. Two users (0 and 10) had motions that differed significantly (but visually imperceptible) from those of other people within the same scenario. There were also the opposite cases (3 and 7), where the model may have lacked features for better discrimination. These results are also highly dependent on the quality of the pre-trained model. For example, results in splits 75 and 80 for User 1, as well as in splits 70 and 80 for User 2, had significant differences in metrics. The adaptive selection of hyperparameters and splits for each unique user is worth paying attention to.

Overall, the proposed user authentication approach shows promising results. As a starting point, this methodology can be used in combination with existing biometric systems to maintain optimal performance in various corner cases. The system can be extremely appealing because it does not need additional hardware. However, the proposed architectures and training methods have a number of drawbacks as far as mobile devices are concerned. As mentioned in the Section~\ref{sec:intro}, biometric systems need to be small and run quickly because of the hardware limitations and usability requirements. The proposed architectures are quite large and the methodology requires customization for each user, which presupposes a built-in backpropagation algorithm. Further research is needed to enhance performance. However we used the industry standardized off-line method to collate with, our results with the on-line approach are promising. We strongly believe that new biometry system will emerge for the new use cases and hardware that is why we share our datasets and code to make new researches possible.
\bibliography{bibliography.bib}
\bibliographystyle{plain}









\appendix
\section{Supplemental Material}
\subsection{Project Code}
The project code can be downloaded via link\footnote{Code: \url{https://github.com/SamsungLabs/MotionID}}. The license is CC BY-NC-SA 4.0. All training details can be found in configs.

\subsection{Datasets}
\subsubsection{Motion Patterns Identification}
The dataset for \emph{Motion Patterns Identification} stage is called \emph{MotionID: IMU all motions}. Due to large size this dataset was divided into three parts (two users per part) which can be downloaded via links\footnote{Part 1: \url{https://www.kaggle.com/datasets/djaarf/motionid-imu-all-motions-part1}}\footnote{Part 2: \url{https://www.kaggle.com/datasets/djaarf/motionid-imu-all-motions-part2}}\footnote{Part 3: \url{https://www.kaggle.com/datasets/djaarf/motionid-imu-all-motions-part3}}.
The license for all three parts is CC BY-NC-SA 4.0.

\subsubsection{User Verification}
The dataset for \emph{User Verification} stage is called \emph{MotionID: IMU specific motion} and can be downloaded via attached link\footnote{\url{https://www.kaggle.com/datasets/djaarf/motionid-imu-specific-motion}}. The license is CC BY-NC-SA 4.0.

Detailed description of datasets can be found via provided links.

\subsection{Computational Resources}
For training of deep learning models we used GPU server with NVIDIA Tesla V100 SXM2 32GB NVLink. Training of one model from \emph{Motion Patterns Identification} stage takes less than 5 minutes. Training of one model from \emph{User Verification} stage takes about 20 hours. Fine-tuning of one model for one user takes about 2 hours.


\end{document}